\documentclass[%
superscriptaddress,
showpacs,preprintnumbers,
nofootinbib,
 amsmath,amssymb,
 aps,
prc
]{revtex4-1}
\usepackage[hidelinks]{hyperref}
\usepackage{float}
\usepackage{graphicx}% Include figure files
\usepackage{dcolumn}% Align table columns on decimal point
\usepackage{bm}% bold math
\usepackage{feynmp-auto}
\usepackage{slashed}
\usepackage[usenames]{xcolor}
\usepackage{appendix}
\usepackage{subcaption}
\newcolumntype{P}[1]{>{\centering\arraybackslash}p{#1}}

\newcommand{\Lagr}{\mathcal{L}}

\newcommand{\mpi}{m_\pi}

\newcommand{\fpi}{f_\pi}
\newcommand{\ChiEFT}{{$\chi$EFT~}} 
\newcommand{\ChiPT}{{$\chi$PT~}} 
\newcommand{\nsat}{n_{\mathrm{sat}}}

\newcommand{\fm}{\,\mathrm{fm}}

\newcommand{\beq}{\begin{equation}}
\newcommand{\eeq}{\end{equation}}
\newcommand{\beqa}{\begin{eqnarray}}
\newcommand{\eeqa}{\end{eqnarray}}

\begin{document}

\preprint{INT-PUB-21-035}

\title{The mass of charged pions in neutron star matter}
\author{Bryce Fore}
\affiliation{Institute for Nuclear Theory and Department of Physics, University of Washington, Seattle,  WA 98195. }
\author{Norbert Kaiser}
\affiliation{Physik-Department T39, Technische Universit\"{a}t M\"{u}nchen, D-85747 Garching, Germany}
\author{Sanjay Reddy}
\affiliation{Institute for Nuclear Theory and Department of Physics, University of Washington, Seattle,  WA 98195. }
\author{Neill C. Warrington}
\affiliation{Institute for Nuclear Theory and Department of Physics, University of Washington, Seattle,  WA 98195. }
\affiliation{Center for Theoretical Physics, Massachusetts Institute of Technology, Cambridge, MA 02139, USA}

\date{ \today}% It is always \today, today,

\begin{abstract}
We examine the behavior of charged pions in neutron-rich matter using heavy-baryon chiral perturbation theory. This study is motivated by the prospect that pions, or pion-like excitations, may be relevant in neutron-rich matter encountered in core-collapse supernovae and neutron star mergers. We find, as previously expected, that the $\pi^-$ mass increases with density and precludes s-wave condensation at $n_B\lesssim \nsat$, where $\nsat \simeq 0.16 \fm^{-3}$ is the nuclear saturation density, and the mass of the $\pi^+$ mode decreases with density. The uncertainty in these predictions increases rapidly for $n_B \gtrsim \nsat $ because low energy constants associated with the two-pion-two-nucleon operators in chiral perturbation theory are poorly constrained. We find that these uncertainties are especially large in symmetric nuclear matter and should be included in the analysis of pion-nucleus interactions at low energy and pionic atoms. In neutron-rich matter, accounting for the self-energy difference between neutrons and protons  related to the nuclear symmetry energy has several effects. It alters the power counting of certain higher-order contributions to the pion self-energy. Previously unimportant but attractive diagrams are enhanced and result in a modest reduction of the pion masses. Furthermore, in the low-wavelength limit, a collective mode with the quantum numbers of the $\pi^+$ appears. 
\end{abstract}

\maketitle
\section{Introduction}
The study of pions in dense nuclear matter has a long history. In  pioneering work in the 1970s, Sawyer and Scalapino \cite{Sawyer:1972,Scalapino:1972,Sawyer&Scalapino:1973} and independently Migdal and collaborators \cite{Migdal:1973,Migdal:1973jkf,Migdal:1974jn,Migdal:1973zm} proposed that pion condensation might lower the free energy density of nuclear matter at high density. In subsequent years, several authors studied its manifestation and implications for nuclei and neutron stars in some detail using models for the pion-nucleon interaction \cite{Baym:1973zk,Baym:1974vzp,Au:1974mbl,Dashen:1974ff,Barshay:1974zw,Weise:1975,Backman:1975,Campbell:1974qt,Campbell:1974qu}. 

Pion condensation occurs at low temperatures when the energy to produce a pion is less than its associated chemical potential. Earlier work has addressed the possibility of condensation of negatively charged pions in neutron stars because the chemical potential for a negative charge, which we denote throughout as $\hat{\mu}$, increases rapidly with density. In the outer core of the neutron star, where the baryon density $n_B \simeq \nsat$ where $\nsat \simeq 0.16 \fm^{-3}$ is the nuclear saturation, $\hat{\mu} \simeq 100$ MeV and could be as large as $300$ MeV in the inner core. Without interactions, a pion condensate with zero-momentum will occur when $\hat{\mu} \ge m_\pi$, where $m_\pi\simeq 140$ MeV is the mass of the pion. However, repulsive s-wave interactions between $\pi^-$ and neutrons increased the pion energy in neutron stars and disfavored condensation of zero-momentum pions. In contrast, a strongly attractive p-wave interaction between pions and nucleons was shown to favor condensation of pions with momentum $k_\pi \simeq \mpi$ and led to a non-uniform ground state \cite{Migdal:1973,Migdal:1974jn,Au:1974mbl}. In the 1980s, more sophisticated but model-dependent analyses, including many-body corrections and correlations between nucleons at short distances, found that even p-wave condensation may not be robust at the densities encountered in neutron stars (for a comprehensive review of these developments see Ref.~\cite{Migdal:1990}). 

In this paper, we revisit calculating the mass of charged pions in dense neutron-rich matter using heavy-baryon chiral perturbation theory (HB$\chi$PT). We do so for the following reasons. First, earlier calculations were based on the mean-field approximation and used simple models for the pion-nucleon interaction, which were poorly constrained by pion-nucleon scattering data. Second, the role of pion coupling to two-nucleon currents has not been studied, and their inclusion is shown to be relevant at the densities of interest. Third, earlier calculations neglected the effect of the nuclear symmetry energy, which induces a large energy self-difference between neutrons and protons in the neutron-rich matter.  Finally, even in the absence of pion condensation, the mass of pionic excitations in the medium is relevant to the description of the ground state and response properties of dense nuclear matter at finite temperatures realized in extreme astrophysical phenomena such as neutron star mergers and core-collapse supernovae \cite{PhysRevC.101.035809}.              

We present results for the pion mass in dense neutron-rich matter using heavy-baryon chiral perturbation theory (HB$\chi$PT) and augment the calculation with a model for the nucleon self-energy. This nucleon self-energy incorporates energy shifts of the neutrons and protons in the dense nuclear medium.  We find that previously neglected two-loop diagrams (involving p-wave interactions in the intermediate state) and the inclusion of in-medium nucleon self-energies lower the pion self-energy relative to previous estimates. In particular, small energy denominators in perturbation theory are produced when the energy difference between neutrons and protons becomes of the order of the pion mass. These small energy denominators promote the importance of certain attractive Feynman diagrams, which lower the $\pi^-$ energy. Modified power counting is corroborated by a recent analysis of experimental data suggesting that the nuclear symmetry energy - the in-medium energy difference between protons and neutrons - can be large at densities of interest to neutron stars. Another important consequence of this energy difference is a negative energy collective mode with the quantum numbers of the $\pi^+$ in the medium.

The calculation presented here improves upon the calculations of the pion self-energy in asymmetric matter using $\chi$PT presented in Refs.~\cite{Kaiser:2001bx,Kolomeitsev:2002gc}. We include all diagrams considered by previous authors while including several others that make a relevant contribution at $n_B\simeq \nsat$ and elucidate the role of the pion coupling to two-nucleon currents. The latter are particularly important in an effective field theory (EFT) approach as they can encode short-distance physics that have been expected to play a role in pion condensation  \cite{Barshay:1974zw}. 
 
The material in this paper is organized as follows. In Section \ref{calculation}, we describe our calculation of the pion self-energy in isospin-asymmetric dense matter. In Section \ref{nucleonnucleon}, we present a parametric model to account for strong nucleon interactions and study its effect on the pion self-energy. In Section \ref{symnuc}, we discuss implications for symmetric nuclear matter, while in Section \ref{results} we discuss neutron-rich matter. Finally, in Section \ref{conclusions}, we offer some conclusions.
\newpage

\section{Calculation}
\label{calculation}
To compute the self-energy of charged pions at non-zero baryon and isospin density, we use heavy-baryon chiral perturbation theory (HB$\chi$PT). This effective field theory (EFT) of mesons and nucleons includes all interactions consistent with the symmetries of QCD and organizes them in a small momentum expansion. We aim to compute the pion self-energy up to $\mathcal{O}(q^6)$, where $q$ an expansion parameter of the EFT. While it is clear which single-nucleon interactions contribute to the pion self-energy up to $\mathcal{O}(q^6)$, it is less clear in the multi-nucleon sector, and we achieve our goal with limited success. In the following, we will take both $m_{\pi}$ and the nucleon Fermi momenta, denoted as $k_f$ to be $\mathcal{O}(q)$.

To calculate the charged pion masses in the medium, we define the self-energy of the \emph{negatively} charged pion at zero momentum, $\Pi(\omega, k_n,k_p)$, through the relation
\begin{equation}
\label{eq:piminus_twopoint}
    \int d^4 x~ e^{i \omega t} \langle T\{\pi^-(x) {\pi^-}^{\dagger}(0) \} \rangle = \frac{i}{\omega^2 - \mpi^2 - \Pi(\omega,k_n,k_p)}~\,,
\end{equation}
where $\langle ..\rangle $ denotes an ensemble average at finite neutron and proton densities characterized by Fermi momenta $k_n$ and $k_p$, respectively.  Isospin symmetry implies  $\Pi(\omega,k_n,k_p) = \Pi(-\omega,k_p,k_n)$. We will calculate $\Pi(\omega, k_n,k_p)$ in diagrammatic perturbation theory and we will see that both one- and two-nucleon operators contribute to the sixth order. We separate these contributions for clarity of presentation: we first present all diagrams up to $\mathcal{O}(q^6)$ generated by single-nucleon operators alone, and then we present the two-nucleon graphs. Single nucleon operators begin contributing to the self-energy at $\mathcal{O}(q^4)$, while two-nucleon operators begin contributing at higher order. The lagrangian of our theory is
\begin{equation}
    \mathcal{L} = \mathcal{L}_{\pi \pi} + \mathcal{L}_{\pi N} + \mathcal{L}_{ \pi N N}
\end{equation}
where $\mathcal{L}_{\pi \pi}$ includes terms with only pion fields, $\mathcal{L}_{\pi N}$ includes single-nucleon terms, and $\mathcal{L}_{\pi N N}$ two-nucleon terms. Each of these terms will be explicitly written below.

Before proceeding to the calculation we discuss some bookkeeping. First, to implement non-zero baryon and isospin density, we use the technique described in \cite{Kaiser:2001bx}; namely, nucleon propagators are
\begin{equation}
\label{eq:nucleon-prop}
    i G_f(p) = \frac{i}{p_0 + i 0^+} - 2\pi \delta(p_0) \theta(k_{f} - |\vec{p}|)
\end{equation}
where $f$ denotes the nucleon species and $k_{f}$ its Fermi momentum. The first term is the vacuum heavy-baryon propagator, while the second term arises from the finite density. Second, we interest ourselves only in zero-momentum pions; operators which produce diagrams that vanish for such kinematics can be discarded. Finally, we neglect all purely pion self-energy graphs in vacuum. These simply renormalize bare parameters to match vacuum properties of the pion and contribute no finite-density information.

\subsection{Single-Nucleon Contributions}
The terms $\mathcal{L}_{\pi\pi}$ and $\mathcal{L}_{\pi N}$ are given by:
\begin{equation}\label{eqn:L0_pipi}
\begin{split}
\Lagr_{\pi\pi} = \, &\frac{\fpi^2}{4}~ {\rm tr}\big( \partial_\mu U \partial^\mu U^\dagger + \chi_+\big) \,,
\end{split}
\end{equation}

\begin{equation}\label{eqn:LO_piN} 
\begin{split}
\Lagr_{\pi N} =\, & \quad \, \bar{N}\big( i v \cdot D + g_A S \cdot u\big) N  \\
& + \bar{N}  \Big( -\frac{i g_A}{2M} \{S\cdot D, v\cdot u\} + c_1 {\rm tr}\chi_+ +\Big(c_2-\frac{g_A^2}{8M}\Big)(v\cdot u)^2+ c_3\, u\cdot u\Big)  N  \\
& + \sum_{i=1}^{23} b_i \bar{N} \mathcal{O}_i N \, ,
\end{split}
\end{equation}
where we have defined the following symbols. $U$ is an $SU(2)$ matrix defined in terms of pion fields $(\pi^+, \pi^0,\pi^-)$ as
\begin{align}
U & = \exp{\left[\frac{i}{f_{\pi}} \begin{pmatrix}
 \pi^0 & \sqrt{2} \pi^+ \\ \sqrt{2} \pi^- & -\pi^0 \end{pmatrix}\right]} \,,
\end{align}
%where $\tr(t^a t^b) = 2 \delta_{ab}$, 
where $\chi_+ = m_{\pi}^2(U + U^{\dagger})$ introduces explicit chiral symmetry breaking (we take $m_{\pi}=139$ MeV), and $f_\pi=92.4$ MeV is the pion decay constant. $N$ is the nucleon field, containing both proton and neutron components, $v_{\mu}=(1,\vec 0)$ is the nucleon four-velocity, $S^\mu=(0, \vec \sigma/2)$ the spin-vector of the nucleon, and $g_A=1.27$ is the axial-vector coupling constant. The chiral covariant derivative and axial-vector quantity are defined in terms of $\xi= \sqrt{U}$ as $D_{\mu}   = \partial_{\mu} + \frac{1}{2} [\xi^{\dagger},\partial_{\mu}\xi]$ and $ u_{\mu}  = i(\xi^{\dagger} \partial_{\mu} \xi - \xi \partial_{\mu} \xi^{\dagger})$. The low-energy constants $c_i, b_i$ and the operators $\mathcal{O}_i$ are described in detail in \cite{Fettes:1998ud}.

The three lines of $\mathcal{L}_{\pi N}$ are, respectively, the leading-order (LO), next-to-leading-order (NLO) and next-to-next-to-leading-order (N2LO) interactions of HB$\chi$PT in the single-nucleon sector. As the lowest order self-energy graphs are $\mathcal{O}(q^4)$, is necessary to consider N2LO interactions to reach $\mathcal{O}(q^6)$. 

%BEGIN FEYNMAN DIAGRAMS

\begin{figure}[h]
\begin{centering}
\begin{fmffile}{first-diagram}
\begin{fmfgraph*}(80,100)\fmfkeep{ld}
   \fmfpen{thick}
   \fmfleft{i}
   \fmfright{o}
   \fmf{dashes,fore=red}{i,v}  
   \fmf{fermion,fore=red}{v,v}
   \fmf{dashes,fore=red}{v,o}
   \fmfdot{v}
   \fmfv{label=LO,fore=red,label.dist=-15,label.angle=90}{v} 
   \end{fmfgraph*} \quad 
\begin{fmfgraph*}(80,100)\fmfkeep{ld}
   \fmfpen{thick}
   \fmfleft{i}
   \fmfright{o}
   \fmf{dashes,fore=blue}{i,v}  
   \fmf{fermion,fore=blue}{v,v}
   \fmf{dashes,fore=blue}{v,o}
   \fmfdot{v}
   \fmfv{label=N2LO,fore=blue,label.dist=-15,label.angle=90}{v}
   \end{fmfgraph*}\quad
\begin{fmfgraph*}(80,100)\fmfkeep{ph}
    \fmfpen{thick}
    \fmfleft{i}
     \fmfright{o}
     \fmf{dashes,fore=blue,tension=2.35}{i,v1}
     \fmf{fermion,fore=blue,left=1}{v1,v2,v1}
     \fmf{dashes,fore=blue,tension=2.35}{v2,o}
     \fmfdot{v1}
     \fmfdot{v2}
     \fmfv{label=NLO,fore=blue,label.dist=-25,label.angle=35}{v1}
     \fmfv{label=NLO,fore=blue,label.dist=-25,label.angle=145}{v2}
 \end{fmfgraph*}\quad
\begin{fmfgraph*}(80,100)\fmfkeep{ds}
   \fmfpen{thick}
    \fmfleft{i}
     \fmfright{o}
     \fmf{dashes,fore=blue,tension=2.5}{i,v1}
     \fmf{fermion,fore=blue,left=1}{v1,v2,v1}
     \fmf{dashes,fore=blue,tension=0.1}{v1,v2}
     \fmf{dashes,fore=blue,tension=2.5}{v2,o}
     \fmfdot{v1}
     \fmfdot{v2}
     \fmfv{label=LO,fore=blue,label.dist=-20,label.angle=35}{v1}
     \fmfv{label=LO,fore=blue,label.dist=-20,label.angle=145}{v2}
   \end{fmfgraph*}\quad \\
\begin{fmfgraph*}(80,100)\fmfkeep{ld}
   \fmfpen{thick}
   \fmfleft{i}
   \fmfright{o}
   \fmf{dashes,fore=green}{i,v}  
   \fmf{fermion,fore=green}{v,v}
   \fmf{dashes,fore=green}{v,o}
   \fmfdot{v}
   \fmfv{label=NLO,fore=green,label.dist=-15,label.angle=90}{v} 
   \end{fmfgraph*} \quad 
\begin{fmfgraph*}(80,100)
%see https://ctan.math.illinois.edu/macros/latex/contrib/feynmf/fmfman.pdf
%page 23 for vertex labelling scheme.
   \fmfpen{thick}
   \fmfsurroundn{v}{8}
   \fmf{dashes,fore=blue,tension=100}{v5,v0}
   \fmf{dashes,fore=blue,tension=100}{v0,v1}
   \fmf{dashes,fore=blue,tension=0}{v2,v4}
   \fmf{fermion,fore=blue,right=0.5,tension=1}{v0,v2}
   \fmf{fermion,fore=blue,right=0.75}{v2,v4}
   \fmf{fermion,fore=blue,right=0.5,tension=1}{v4,v0}
   \fmfdot{v0}
   \fmfdot{v[2]}
   \fmfdot{v[4]}
   \fmfv{label=LO,fore=blue,label.dist=5,label.angle=-90}{v0}
   \fmfv{label=LO,fore=blue,label.dist=5,label.angle=0}{v2}
   \fmfv{label=LO,fore=blue,label.dist=5,label.angle=180}{v4}
\end{fmfgraph*} \quad
\begin{fmfgraph*}(80,100)\fmfkeep{three-pion}
    \fmfpen{thick}
    \fmfleft{i}
    \fmfright{o}
    \fmftop{t1,t2,t3}
    \fmf{dashes,fore=blue}{i,v}
    \fmf{dashes,fore=blue}{v,o}
    \fmf{phantom,tension=0}{v,t2}
    \fmffreeze
    \fmf{fermion,fore=blue,right}{v,t2,v}
    \fmf{dashes,fore=blue}{v,t2}
    \fmfdot{v}
    \fmfdot{t2}
    \fmfv{label=LO,fore=blue,label.dist=5,label.angle=-90}{v}
    \fmfv{label=LO,fore=blue,label.dist=5,label.angle=90}{t2}
\end{fmfgraph*} \quad
\begin{fmfgraph*}(80,100)\fmfkeep{four-pion}
    \fmfpen{thick}
    \fmfleft{i}
    \fmfright{o}
    \fmftop{t1,t2,t3,t4}
    \fmf{dashes,fore=blue,tension=100}{i,v}
    \fmf{dashes,fore=blue,tension=100}{v,o}
    \fmf{dashes,fore=blue}{v,t2}
    \fmf{dashes,fore=blue}{v,t3}
    \fmf{fermion,fore=blue,left}{t2,t3,t2}
    \fmfdot{v}
    \fmfdot{t2}
    \fmfdot{t3}
    \fmfv{label=LO,fore=blue,label.dist=5,label.angle=-90}{v}
    \fmfv{label=LO,fore=blue,label.dist=5,label.angle=180}{t2}
    \fmfv{label=LO,fore=blue,label.dist=5,label.angle=0}{t3}
\end{fmfgraph*} \quad
\end{fmffile}
\end{centering}
\caption{Feynman diagrams contributing to the \emph{negative} pion self-energy. Solid and dashed lines represent nucleons and pions, respectively, and vertices are labeled by their chiral order. Red, green and blue diagrams are \textcolor{red}{$\mathcal{O}(q^4)$}, \textcolor{green}{$\mathcal{O}(q^5)$}, \textcolor{blue}{$\mathcal{O}(q^6)$}.}
\label{fig:single-nucleon-diagrams}
\end{figure}
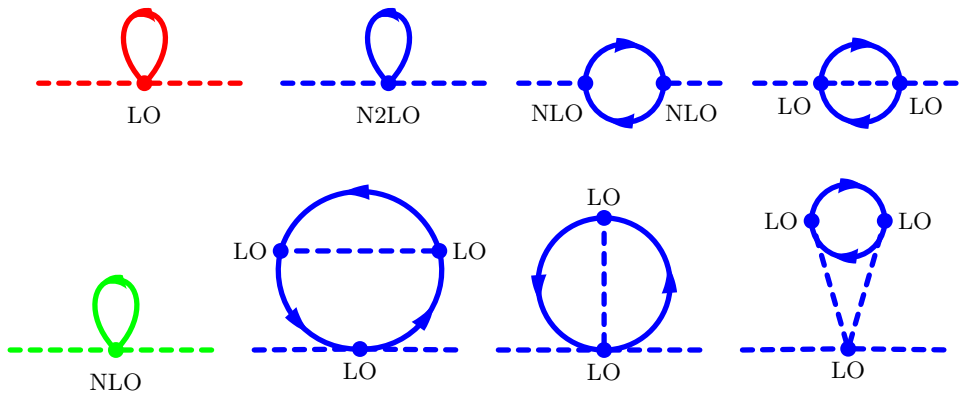

%END FEYNMAN DIAGRAMS

All diagrams up to $\mathcal{O}(q^6)$ arising from single-nucleon interactions are shown in Fig. \ref{fig:single-nucleon-diagrams}. Dotted lines denote pions while solid arrows denote nucleons. The labels on each vertex (LO, NLO, etc.) denote any interaction from $\mathcal{L}_{\pi N}$ at the order indicated. The zero-momentum of the pion restricts the number of contributing graphs; for example any graph where an external pion connects to a leading-order axial coupling vanishes.

Since the nucleon propagator can be separated into a sum of vacuum and finite density pieces, each diagram in Fig. \ref{fig:single-nucleon-diagrams} can be decomposed into a sum of terms with an increasing number of finite-density insertions: from zero to the number of nucleon lines. Contributions with zero finite-density insertions either vanish or renormalize vacuum parameters; single finite density insertions produce terms linear in the density; two or more finite density insertions produce higher (often fractional) powers of the density. Because any given diagram may contribute at several orders in the density, we combine contributions from various diagrams in a way most useful for us. We attempt, however, to make the accounting of all terms clear. 

We denote the sum of all diagrams in Fig. 1 as $\Pi_{\pi N}$, and it is given by:
\begin{equation}
    \Pi_{\pi N}(\omega,k_n,k_p) = \Pi_{ld}(\omega,k_n,k_p) + \Pi_{ds}(\omega,k_n,k_p) + \Pi_{ph}(\omega,k_n,k_p)+\Pi_{cor}(\omega,k_n,k_p)~.
\end{equation}
The first term, $\Pi_{ld}$, gathers all contributions linear in the density and equals
\begin{equation}
\Pi_{ld}(\omega,k_n,k_p)=T^-(\omega)~\frac{k_n^3-k_p^3}{3\pi^2} - T^+(\omega)~\frac{k_n^3+k_p^3}{3\pi^2}\,.
\end{equation}
where $k_n, k_p$ are the neutron and proton Fermi momenta. Within $\Pi_{ld}$ are all $\mathcal{O}(q^4)$ and $\mathcal{O}(q^5)$ diagrams, as well as single density insertions from all $\mathcal{O}(q^6)$ diagrams except the particle-hole diagram (which goes as $k_f^5$). Interestingly, higher density insertions of the fourth diagram vanish due to the appearance of a $\delta'(p_0)$ in the momentum integral \cite{Kolomeitsev:2002gc}. $T^+(\omega)$ and $T^-(\omega)$ are respectively the isoscalar and isovector scattering amplitudes, and the diagrams summed in $\Pi_{ld}$ build up these scattering amplitudes to N2LO. 

The real part of the isoscalar amplitude is\footnote{In this work we focus solely on properties of the pion obtainable from the real-part of the self-energy.}
\begin{equation}\label{eq:Tplus}
T^+(\omega)=\frac{\sigma_N-\beta \omega^2}{f_\pi^2}+\frac{3 g_A^2\mpi^3}{16\pi f_\pi^4} + \frac{3 g_A^2\mpi Q^2 \zeta}{64\pi f_\pi^4} \,,
\end{equation}
where $\sigma_N=-4 c_1\mpi^2-9g_A^2\mpi^3/64\pi\fpi^2$ is the pion-nucleon sigma term \cite{Sainio:2001bq} and $\beta=-2(c_2+c_3)+g_A^2/4M$. To fix low-energy constants we adopt the same strategy as Ref.~\cite{Kolomeitsev:2002gc}, namely $c_1$ is chosen to reproduce $\sigma_N \simeq 45 \pm 15 $ MeV, the range probed by phenomenological and lattice calculations \cite{Hoferichter:2015dsa, Borsanyi:2020bpd}, and the combination $c_2+c_3$ is tuned to obtain the empirical value $T^+(\mpi)\simeq 0$. Finally, the parameter $\zeta$ reflects freedom of choice in the interpolating field for the pion \cite{Park:2001ht,Meissner:2001gz}; as in Ref.~\cite{Kolomeitsev:2002gc} we set $\zeta = 0$, a condition obtained by requiring the residue of the propagator at the pion pole be equal to one.

The real-part of the isovector amplitude is 
\begin{equation}
\label{eq:Tminus}
T^-(\omega)=\frac{\omega}{2f_\pi^2}+\frac{\gamma \omega^3}{8\pi^2 f_\pi^4} - \frac{ \omega^2 Q}{8\pi^2 f_\pi^4}\ln{ \frac{\lvert \omega+Q \rvert}{m_\pi} } \,,
\end{equation}
where $Q=\sqrt{\omega^2 - m_{\pi}^2}$ and $\gamma = (g_A \pi f_{\pi}/M)^2+\text{ln}(2\Lambda/m_{\pi})$ \cite{Kolomeitsev:2002gc}. The first term is due to the leading-order Weinberg-Tomozawa interaction \cite{Weinberg:1966kf,Tomozawa:1966jm}, while subsequent terms are N2LO corrections. It will be noticed that none of the $b_i$ coefficients explicitly appear in $T^-$. A part of their effect is incorporated in the tuning of $\Lambda$ such that the empirical value of $T^-(m_{\pi}) = 1.85$ fm is obtained. The residual $\mathcal{O}(\omega^3)$ dependence these operators introduce into $T^-$ have little effect on the outcome of the calculation and are ignored. 

Next, $\Pi_{ds}$ comes from the double-scattering diagram with two density insertions and is given by
\begin{equation}
\Pi_{ds}(\omega,k_n,k_p) =
\frac{\omega^2}{3(4\pi \fpi)^4} \Big\{L(\omega;k_n,k_n)+L(\omega;k_p,k_p)+2L(\omega;k_n,k_p)\Big\}\,,
\end{equation}
with the logarithmic function
\beq
\begin{split}
L(\omega; k_n,k_p) = & \,\,4 k_n k_p (3k_n^2 +3k_p^2+Q^2) + 8 Q (k_n^3-k_p^3)\ln{\frac{| Q+k_n-k_p |}{|Q-k_n+k_p|}}- 8 Q (k_n^3+k_p^3)\ln{\frac{Q+k_n+k_p}{|Q-k_n-k_p|}} \\
&+ \Big[3(k_n^2-k_p^2)^2 + 6 Q^2 (k_n^2+k_p^2)-Q^4\Big]\ln{\frac{|(k_n-k_p)^2-Q^2|}{|(k_n+k_p)^2-Q^2|}}\,, \text{ when }\omega^2 - m_{\pi}^2 > 0 \\
= & \,\, 4 k_n k_p (3k_n^2 +3k_p^2-q^2) + 16 q \Big((k_n^3 - k_p^3)\text{arctan}\frac{k_n-k_p}{q} - (k_n^3 + k_p^3)\text{arctan} \frac{k_n+k_p}{q} \Big) \\
& + \Big[ 3(k_n^2 - k_p^2)^2 - 6 q^2 (k_n^2 + k_p^2) - q^4 \Big] \ln{\frac{(k_n-k_p)^2+q^2}{(k_n+k_p)^2+q^2}}\,, \text{ when }\omega^2 - m_{\pi}^2 < 0 ~.
\end{split}
\eeq
with $q = \sqrt{m_\pi^2-\omega^2}$.

The ``particle-hole" diagram $\Pi_{ph}$, with two NLO axial-vector interactions, equals
\begin{equation}
\Pi_{ph} (\omega,k_n,k_p)=
\frac{g_A^2 \omega }{f_\pi^2} \bigg(\frac{k_p^5-k_n^5}{10\pi^2 M^2}\bigg)\,.
\label{eqn:pi_ph}
\end{equation}
Only the single-density insertions of this diagram are non-zero. This fact, combined with the NLO axial-vector interaction coupling to the nucleon (rather than the pion) momentum, produces the $k_f^5$ dependence. This diagram is suppressed in systems with small isospin asymmetry and is therefore unimportant in the analysis of pionic atoms \cite{Kolomeitsev:2002gc}. In contrast, it is non-negligible in isospin asymmetric environments like neutron stars.   

Finally, the two-density insertion contributions of the final two $\mathcal{O}(q^6)$ diagrams combine to give
\begin{equation}
    \Pi_{cor}(\omega,k_n,k_p) = \frac{g_A^2}{20(4\pi f_{\pi})^4}\bigg( \big\{Q^2(\zeta + 2)+5 m_{\pi}^2 \big\}[H(k_p,k_p)+H(k_n,k_n)]+Q^2(8\zeta - 4)H(k_p,k_n)\bigg)~,
\end{equation}
where the $H$ function is \cite{Kolomeitsev:2002gc}:
\begin{equation}
\begin{split}
    H(k_p,k_n) = & \, 8 k_p k_n(m_{\pi}^2-k_p^2-k_n^2) 
     + 16 m_{\pi} (k_p^3+k_n^3)\text{arctan}\frac{k_p+k_n}{m_{\pi}}-16 m_{\pi} |k_p^3-k_n^3|\text{arctan}\frac{|k_p-k_n|}{m_{\pi}} \\
    & + 2\bigg[(k_p^2-k_n^2)^2 - 4 m_{\pi}^2(k_p^2+k_n^2)-m_{\pi}^4\bigg]\ln  \frac{m_{\pi}^2+(k_p+k_n)^2}{m_{\pi}^2+(k_p-k_n)^2} ~.
\end{split}
\end{equation}
As previously pointed out, $\Pi_{cor}$ is numerically small \cite{Kolomeitsev:2002gc}.
This concludes a full accounting of the finite-density contributions to the pion self-energy  generated by single-nucleon interactions up to $\mathcal{O}(q^6)$. We note that such an account was already accomplished by one of us in \cite{Kaiser:2001bx,Kolomeitsev:2002gc} (a similar calculation is presented in \cite{Park:2001ht}). We have included this known information because it will be useful later and to demarcate between our work and others. The new contributions we include are described in the following section.

\subsection{Two-Nucleon Contributions}
\label{subsec:2nucleon} 
The systematic inclusion of nucleon-nucleon (NN) interactions presents several challenges as the unnaturally large NN scattering lengths require a non-perturbative approach. Despite significant effort, even in the vacuum, a systematic EFT framework to describe NN interactions, including pions, remains elusive. In an approach pioneered by Weinberg, called Chiral EFT, one derives a potential by systematically including the contributions of pion loops and associated contact interactions (see Ref.~\cite{Epelbaum:2008ga} for a review). This potential is then employed in the Schrodinger equation to include non-perturbative effects. While this approach has been phenomenologically successful, it relies on a fine-tuned range of values for the UV cutoff, which obscures systematic power counting \cite{Hammer:2019poc}. On the other hand, a modified power-counting scheme that preserves renormalization group invariance developed by Kaplan, Savage, and Wise, which includes short-distance physics non-perturbatively and pions perturbatively, works well in some partial waves but fails to converge in others \cite{Hammer:2019poc}.

Coupling to a finite density of nucleons complicates matters further, as it introduces $k_F$ as an additional dimensionful scale 
which is not small ($k_F \sim 300$ MeV at nuclear saturation density, $n_{\text{sat}} = 0.16 \, \text{fm}^{-3}$).
Furthermore, as already pointed out by Weinberg, nucleon propagators can scale as $\mathcal{O}(q^{-1})$ or $\mathcal{O}(q^{-2})$ depending on kinematics \cite{WEINBERG1990288,WEINBERG19913}. This complicates power counting: in the first case, NN interactions begin contributing to the pion self-energy at $\mathcal{O}(q^6)$, while in the second at $\mathcal{O}(q^5)$. An in-medium power counting scheme where all nucleon propagators are counted as $\mathcal{O}(q^{-2})$ was proposed in \cite{Oller_2009}. This scheme requires a non-perturbative resummation of NN interactions, and it was found that all $\mathcal{O}(q^{5})$ diagrams cancel. This suggests that NN interactions begin to contribute to the pion self-energy at $\mathcal{O}(q^6)$, which is why we face this problem in this work.

Absent a clear optimal scheme for power-counting NN interactions, we model them instead. The main ingredient of our model is a nucleon self-energy, fit to dense matter data, that dresses nucleon lines. To estimate uncertainties, we include the effects of the lowest-order $\pi \pi NNNN$ operators. These multi-nucleon operators have unknown coefficients, which may be large, and it will be seen that they can render the pion mass quite uncertain. We further estimate uncertainties by computing several higher-loop diagrams. 

The leading order contribution of NN interactions in many-body perturbation theory is obtained by replacing the free nucleon propagator in Eq.~\ref{eq:nucleon-prop} by a dressed, in-medium nucleon propagator of the form 
\begin{equation}
    i G_f(p) = \frac{i}{p_0 - \Sigma_f+i0^+}-2\pi\delta(p_0 - \Sigma_f) \theta(k_{f}-|\vec{p}|)~\,, 
\label{eq:nucleon-mfprop}
\end{equation}
where $\Sigma_f$ is the self-energy of the nucleon with isospin label $f$. This can be seen by inserting into nucleon lines a leading-order NN contact interaction
\begin{equation}
     \delta \mathcal{L} = -C_S (\bar{N}N)^2 - C_T(\bar{N} \vec{\sigma} N)^2\,.
\label{eq:NN_LO}
\end{equation}
To be clear, this is our model: we modify nucleon propagators as specified in Eq. \ref{eq:nucleon-mfprop}, then recompute all diagrams from the previous section. Many diagrams are unaffected by this change, for example those that contribute to $\Pi_{ld}$ and $\Pi_{ds}$. Others, however, do change. In particular, diagrams that contain a particle-hole intermediate state, such as $\Pi_{ph}$, are affected by the dressing of nucleon propagators. They are parametrically enhanced in asymmetric matter where $\Sigma_n \neq  \Sigma_p$.  When the difference $\Sigma_n - \Sigma_p \simeq \mathcal{O}(m_{\pi})$, such diagrams can be promoted one order in the low-momentum expansion. Exactly how the promotion occurs will be explained later; however, a simple pattern emerges: the net effect of NN interactions is to multiply diagrams with particle-hole intermediate states by the dimensionless factor
\begin{equation}
\xi(\omega) = \frac{\omega}{\omega-\big( \Sigma_n - \Sigma_p \big) }\,.   
\label{eq:enchancement}
\end{equation}

\begin{figure}[t]
\begin{centering}
\begin{fmffile}{second-set-of-diagrams}
\begin{fmfgraph*}(120,80)\fmfkeep{pw}
   \fmfpen{thick}
    \fmfleft{i}
     \fmfright{o}
     \fmf{dashes,tension=2.85}{i,v1}
     \fmf{fermion,left=1}{v1,v2}
     \fmf{fermion,left=1}{v2,v3}
     \fmf{dashes,tension=0.}{v1,v2}
     \fmf{fermion,left=1}{v3,v1}
     \fmf{dashes,tension=2.85}{v3,o}
     \fmfdot{v1}
     \fmfdot{v2}
     \fmfdot{v3}
     \fmfv{label=LO,label.dist=5,label.angle=-130}{v1}
     \fmfv{label=LO,label.dist=5,label.angle=0}{v2}
     \fmfv{label=NLO,label.dist=5,label.angle=-50}{v3}
   \end{fmfgraph*}\quad 
\end{fmffile}
\end{centering}
\caption{A diagram that is nominally $\mathcal{O}(q^7)$, that can be promoted to order $\mathcal{O}(q^6)$ in asymmetric matter.}
\label{fig:pw-diagram}
\end{figure}
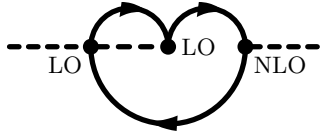

Since diagrams may be promoted, $\mathcal{O}(q^7)$ graphs must be considered. While it is beyond the scope of this work to enumerate all $\mathcal{O}(q^7)$ graphs, we can enumerate the subset generated solely by pion-single-nucleon interactions. We find a single promoted $\mathcal{O}(q^7)$ graph, depicted in Fig. \ref{fig:pw-diagram}, 
This graph includes the leading-order attractive p-wave pion-nucleon interaction in the intermediate state. In Fig.~\ref{fig:pw-diagram}, we show only one intermediate state. Others are obtained by connecting the p-wave vertex to the lower Fermion line, by switching the location of the LO two-pion vertex, and by setting the intermediate pion to be either charged or neutral. We denote the sum of these diagrams $\Pi_{pw}$, which is equal to
\begin{align}
\Pi_{pw} (\omega,k_n,k_p) = \frac{g_A^2\omega }{4M f_\pi^4} \Big\{  2I(\omega;k_p,k_p)-2I(\omega;k_n,k_n) -2 K(\omega;k_n,k_p) + I(0;k_p,k_p)-I(0;k_n,k_n) - K(0;k_n,k_p) \Big\} \,.
\label{eqn:pi_pw_wt}
\end{align}
The functions $I(\omega; k_n,k_p)$ and $K(\omega; k_n,k_p)$ arise from the following principal-value integrals of a pion propagator over two Fermi spheres:
\begin{align}
    I(\omega;k_n,k_p) & = \mathcal{P} \int \frac{d^3l_1 d^3l_2}{(2\pi)^6}\theta\big(k_n-|\vec l_1|\big)\theta\big(k_p-|\vec l_2|\big)\frac{(\vec l_1-\vec l_2)^2}{(\vec l_1-\vec l_2)^2+m_{\pi}^2-\omega^2}\,, \nonumber \\
    K(\omega;k_n,k_p) & = \mathcal{P} \int \frac{d^3l_1 d^3l_2}{(2\pi)^6}\theta\big(k_n-|\vec l_1|\big)\theta\big(k_p-|\vec l_2|\big)\frac{2({\vec l_1}^{\,\,2}-{\vec l_2}^{\,\,2})}{(\vec l_1-\vec l_2)^2+m_{\pi}^2-\omega^2} \,. 
\end{align}
We provide explicit analytical expressions for these functions in the appendix. 

\begin{figure}[b]
\begin{centering}
\begin{fmffile}{multi-nucleon-diagrams}
 \begin{fmfgraph*}(120,80)\fmfkeep{cd}
 \fmfleft{i}
     \fmfright{o}
    \fmfpen{thick}
    \fmftop{t0,t1,t2,t3}
        \fmfbottom{b0,b1,b2,b3}
        \fmf{phantom}{t1,v1,b1}
        \fmf{phantom}{t2,v2,b2}
        \fmffreeze
     \fmf{dashes,tension=3}{i,v1}
     \fmf{fermion,left=1}{v1,v2,v1}
     \fmf{fermion,left=1}{v1,v1}
     \fmf{dashes,tension=3}{v2,o}
     \fmfdot{v1}
     \fmfdot{v2}
     \fmfv{label=$c_D$,label.dist=5,label.angle=-120}{v1}
     \fmfv{label=NLO,label.dist=6,label.angle=-50}{v2}
     \fmfv{label=(a),label.dist=-20,label.angle=0}{b2}
 \end{fmfgraph*}
  \begin{fmfgraph*}(80,60)\fmfkeep{d2}
   \fmfpen{thick}
   \fmftop{t0,t1,t2}
        \fmfbottom{b0,b1,b2}
        \fmf{phantom}{t1,v1,b1}
        \fmf{phantom}{t2,v2,b2}
        \fmffreeze
   \fmfleft{i}
   \fmfright{o}
   \fmf{dashes}{i,v}  
   \fmf{fermion,tension=.8,right=5}{v,v}
   \fmf{fermion,tension=.45,right=5}{v,v}
   \fmf{dashes}{v,o}
\fmfv{decor.shape=triangle,decor.filled=shaded,decor.size=12,decor.angle=180,label=$X_2$,label.dist=10,label.angle=-90}{v}
\fmfv{label=(b),label.dist=0,label.angle=0}{b1}
 \end{fmfgraph*}
\end{fmffile}
\end{centering}
\caption{Two-nucleon contributions to the pion self-energy. The label $X_2$ denotes any of $D_2,D_2',E_2,E_2'$.}
\label{fig:multi-nucleon-diagrams}
\end{figure}
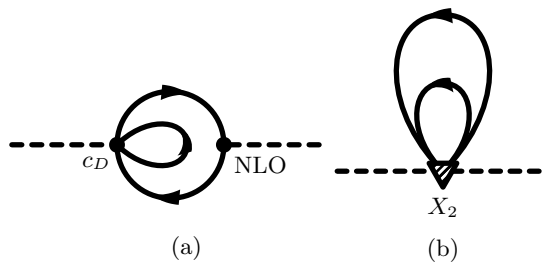

As previously stated, we estimate the uncertainties of our model by examining the contributions from operators that involve pion couplings to two nucleons. These are certainly neglected in our model of NN interactions because they involve external pion lines . The ``$c_D$ operator", which is well-studied in the context of three-nucleon forces and is described by   
\begin{equation}
\mathcal{\delta L} = \frac{c_D}{2M \Lambda f_{\pi}^2}  \bar{N} \{S\cdot D, v\cdot u\}  N ~\bar{N}N
\end{equation}
produces  diagram (a) in Fig. \ref{fig:multi-nucleon-diagrams}. If $c_D \sim \mathcal{O}(1)$, then the $c_D$ diagram is nominally $\mathcal{O}(q^9)$. Since this diagram contains a particle-hole loop, it can be enhanced in asymmetric matter for the reason discussed earlier. Despite the promotion, this diagram is still small at $n_B \lesssim 2\nsat$, and we  neglect it. The remaining lowest-order two-nucleon operators are
%\begin{align}
%\delta \mathcal{L} = & - D_2 m_{\pi}^2 \big(N^T P_a N\big)^{\dagger} \big(N^T P_a N\big) - D_2' m_{\pi}^2 \big(N^T P_a' N\big)^{\dagger} \big(N^T P_a' N\big) \nonumber \\
%& +\frac{D_2 m_{\pi}^2}{2} \frac{\pi^2}{f_{\pi}^2 }\big(N^T P_a N\big)^{\dagger} \big(N^T P_a N\big) +\frac{D_2' m_{\pi}^2}{2} \frac{\pi^2}{f_{\pi}^2 }\big(N^T P_a' N\big)^{\dagger} \big(N^T P_a' N\big) \nonumber \\
%& + \frac{E_2}{2} \frac{\partial_0 \pi \partial_0 \pi}{f_{\pi}^2} \big(N^T P_a N\big)^{\dagger} \big(N^T P_a N\big) + \frac{E_2'}{2} \frac{\partial_0 \pi \partial_0 \pi}{f_{\pi}^2} \big(N^T P'_a N\big)^{\dagger} \big(N^T P'_a N\big) 
%\end{align}
\begin{align}
\delta \mathcal{L} = & - D_2 m_{\pi}^2\big(1-\frac{\vec{\pi}^2}{2 \fpi^2}\big) \big(N^T P_a N\big)^{\dagger} \big(N^T P_a N\big) - D_2' m_{\pi}^2\big(1-\frac{\vec{\pi}^2}{2 \fpi^2}\big) \big(N^T P_a' N\big)^{\dagger} \big(N^T P_a' N\big) \nonumber \\
& + (\partial_0 \pi)^2~\left(\frac{E_2}{2f_{\pi}^2}   \big(N^T P_a N\big)^{\dagger} \big(N^T P_a N\big) + \frac{E_2'}{2f_{\pi}^2} \big(N^T P'_a N\big)^{\dagger} \big(N^T P'_a N\big) \right)
\end{align}
where $P_a = \frac{1}{\sqrt{8}} \tau_2 \tau_a \sigma_2$ and  $P'_a = \frac{1}{\sqrt{8}} \tau_2 \sigma_a \sigma_2$ project onto the $(S,I)=(0,1)$ and $(S,I)=(1,0)$ channels respectively, where $S$ is spin and $I$ is isospin.
Note that we have only written operators that contribute to the self-energy of a stationary pion, and for ease, we have switched to non-relativistic notation. The latter is simply a convenience and does not change the final result. The operators in the first row renormalize the ${}^1 S_0$ and ${}^3 S_1$ couplings and are, in principle, accounted for in our model self-energy. The following two rows of operators include explicit pion fields and produce new contributions to the pion self-energy. Their corresponding Feynman diagrams are shown in (b) of Fig. \ref{fig:multi-nucleon-diagrams} and contributions are:
\begin{align}
    \Pi_{D_2}(\omega,k_n,k_p) & = -\frac{D_2 m_{\pi}^2}{4 f_{\pi}^2}\big(n_n^2 + n_p^2 + n_n n_p\big) \nonumber \\
    \Pi_{E_2}(\omega,k_n,k_p) & = - \frac{E_2 \omega^2 }{4 f_{\pi}^2} \big(n_n^2 + n_p^2 + n_n n_p\big) \nonumber \\
    \Pi_{D'_2}(\omega,k_n,k_p) & = -\frac{3 D'_2 m_{\pi}^2}{4 f_{\pi}^2} n_n n_p \nonumber \\
    \Pi_{E'_2}(\omega,k_n,k_p) & = - \frac{3 E'_2 \omega^2}{4 f_{\pi}^2} n_p n_n~.
\end{align}
with proton and neutron densities $n_p = k_p^3/3\pi^2, n_n = k_n^3/3\pi^2$. We parameterize the couplings above in the following way
\begin{equation}
\Big(D_2,E_2,D_2', E_2'\Big) = \frac{4 \pi }{M \mu}\frac{1}{\Lambda_{NN} \mu} \Big(d_2(\mu), e_2(\mu), d_2'(\mu), e_2'(\mu) \Big)~,
\end{equation}
where $d_2,e_2,d_2',e_2'$ are dimensionless, $\mu = m_{\pi}$ and $\Lambda_{NN} = (16 \pi f_{\pi}^2/g_A^2 M) \simeq 300$ MeV 
%\footnote{Note that our $f_{\pi} = f_{\pi}^{KSW}/\sqrt{2}$; our definitions for $\Lambda_{NN}$ are the same \cite{Kaplan:1998we}.}. 
We denote any of the four coefficients above as $X_2(\mu) = (4 \pi x_2(\mu)/M \mu^2 \Lambda_{NN})$ as needed for brevity. As with any effective field theory, the size of these bare coefficients depends on both the chosen renormalization scheme and mass scale $\mu$. In an EFT framework, the $\mu$ dependence of $x_2$ is related to the nucleon-nucleon interactions and must be evolved consistently. However, since we employ a phenomenological model for NN interactions, we explore a range of values for $x_2$, which is motivated below to assess the natural size of the potential contributions from these two-nucleon operators.

First, we discuss $x_2(\mu)$ within KSW's power counting scheme \cite{Kaplan:1998we}. This a more appropriate choice since we are considering single perturbative insertions of the $X_2$ operators into the pion self-energy. 
%We choose this particular renormalization scheme over Weinberg's because we are considering single perturbative insertions of the $X_2$ operators into the pion self-energy; this is consistent with KSW power counting and not Weinberg. and will ultimately treat them as unknown parameters that we vary over a reasonable range ($x_2(\mu=m_{\pi}) \sim \mathcal{O}(1)$ within KSW power counting). 
In the KSW approach the $D_2$ operator is needed renormalize the $^1S_0$ two-nucleon scattering amplitude \cite{Kaplan:1998we} \footnote{We thank Emanuele Mereghetti for alerting us to its importance.}. However, its value at a specific renormalization scale has not yet been extracted from nucleon-nucleon scattering data. This is because $D_2$ makes its first appearance only the combination $C^{({}^1S_0)} + \mpi^2 D_2$, and can only be separated from $C^{({}^1 S_0)}$ in a high order calculation. Ref.~\cite{Beane:2002xf} suggests that $|D_2(\mu) \mpi^2| \lesssim \eta |C_S(\mu)|$, where $ 1/15< \eta <1/5$ for $3\mpi >\mu> \mpi$ but posits that it could be larger. In principle, EFT analysis of  pion-nucleus scattering could provide useful constraints \cite{Hammer:2019poc}, or lattice QCD calculations of the quark mass dependence of the two-nucleon scattering amplitude can provide more realistic constraints in the future \cite{PhysRevC.86.054001,Beane:2002xf,PhysRevC.85.044001,PhysRevLett.97.012001}. In a  calculation with heavier than physical pion masses, it was found that $| \mpi^2 D_2(\mu)/C_0(\mu) | \sim 0.1$ at $\mu = 350$ MeV \cite{PhysRevLett.97.012001}. This implies $d_2(\mu) \sim \mathcal{O}(1)$. For now, we shall assume that $d_2(\mu)$ is unconstrained by data and explore the range $|d_2(\mu)| \le 1$. 

Comparatively fewer constraints appear in the literature for $d_2', e_2'$ and $e_2$. A calculation of the pion-deuteron scattering length using KSW power counting constrains the linear combination $D_2' + E_2'$ \cite{borasoy2002s}. The rather uncertain isoscalar pion-nucleon scattering length is needed to determine these coefficients, and the window $-5 \leq d_2' + e_2' \leq 5$ is found. Absent constraints on the coefficients individually, we assume $|d_2'| \leq 1$ and $|e_2'| \leq 1$ in our analysis. We find no constraints for $e_2$ and similarly assume $|e_2| \leq 1$. 

We conclude our analysis of the $X_2$ operators by discussing their size in Weinberg power counting. Within this scheme, the $X_2$ coefficients are expected to scale as $X_2(\Lambda_b) \sim (4 \pi \eta(\Lambda_b))/(f_{\pi}^2 \Lambda_b^2)$, where the dimensionless $\eta(\Lambda_b)$ are expected to be $\mathcal{O}(1)$ at $\Lambda_b = 650$ MeV\footnote{We thank Evgeny Epelbaum for discussions on this point.}. One finds the numerical value of $X_2(\Lambda_b)$ estimated in the Weinberg scheme to be within a factor of two of the $X_2(\mu)$ estimated in the KSW scheme. The error bands in the plots to follow, therefore, represent the uncertainty produced by perturbative insertions of the $X_2$ operators within both Weinberg and KSW power counting.
 
\begin{figure}[t]
\begin{centering}
\begin{fmffile}{third-set-of-diagrams}
\begin{fmfgraph*}(120,80)\fmfkeep{cd}
 \fmfleft{i}
     \fmfright{o}
    \fmfpen{thick}
    \fmftop{t0,t1,t2,t3,t4}
        \fmfbottom{b0,b1,b2,b3,b4}
        \fmf{phantom}{t1,v,b1}
        \fmf{phantom}{t2,v1,b2}
        \fmf{phantom}{t3,v2,b3}
        \fmffreeze
     \fmf{dashes,tension=3}{i,v}
     \fmf{fermion,left=1}{v,v1}
     \fmf{fermion,left=1}{v1,v}
     \fmf{fermion,left=1}{v1,v2}
     \fmf{fermion,left=1}{v2,v1}
     \fmf{dashes,tension=3}{v2,o}
     \fmfdot{v}
     \fmfdot{v2}
     \fmfv{label=NLO,label.dist=5,label.angle=-120}{v}
     \fmfv{label=NLO,label.dist=6,label.angle=-50}{v2}
     \fmfv{label=(a),label.dist=0,label.angle=0}{b2}
     \fmfv{decor.shape=circle,decor.filled=shaded, decor.size=10,label=V$_0$,label.dist=-24,label.angle=90}{v1}
 \end{fmfgraph*}
 \quad \qquad
\begin{fmfgraph*}(80,60)\fmfkeep{c02}
 \fmfleft{i}
     \fmfright{o}
    \fmfpen{thick}
    \fmftop{t0,t1,t2,t3,t4}
        \fmfbottom{b0,b1,b2,b3,b4}
        \fmf{phantom}{t1,v1,b1}
        \fmf{phantom}{t2,v2,b2}
        \fmffreeze
     \fmf{dashes,tension=3}{i,v2}
     \fmf{fermion,top=1}{t1,v2}
     \fmf{fermion,top=1}{t1,t3}
     \fmf{fermion,top=1}{v2,t3}
     \fmf{fermion,tension=.1,right=1}{t3,t1}
     \fmf{fermion,tension=50,right=1.5}{t3,t1}
     \fmf{dashes,tension=3}{v2,o}
     \fmfv{decor.shape=square,decor.filled=shaded, decor.size=6,label=LO/NLO,label.dist=-15,label.angle=90}{v2}
     \fmfv{decor.shape=circle,decor.filled=shaded, decor.size=10,label=V$_0$,label.dist=-18,label.angle=0}{t1}
     \fmfv{decor.shape=circle,decor.filled=shaded, decor.size=10,label=V$_0$,label.dist=8,label.angle=0}{t3}
     \fmfv{label=(b),label.dist=0,label.angle=0}{b2}
 \end{fmfgraph*}\quad
\end{fmffile}
\end{centering}
\caption{Additional graphs produced by NN interactions. Here $V_0$ denotes $C_S$ or $C_T$ insertions.}
\label{fig:higher-order-diagrams}
\end{figure}
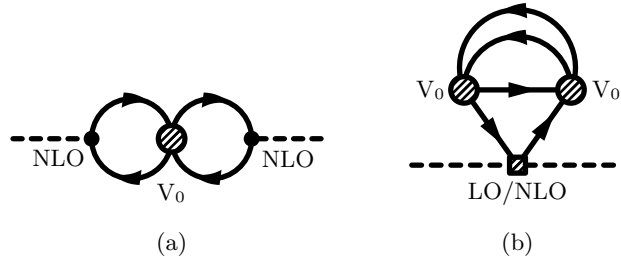

The leading diagrams that include NN interactions beyond the mean field approximation are shown in Fig.~\ref{fig:higher-order-diagrams}. The left diagram is a vertex correction produced by leading-order NN interactions. While analogous diagrams are ultimately responsible for suppressing p-wave condensation, the left diagram of Fig.~\ref{fig:higher-order-diagrams} vanishes for a zero-momentum pion because it is proportional to the  nucleon velocity, which averages to zero in a rotationally symmetric medium. 
%\NW{I show this on p. 489 of my calculation notebook.}
To assess the importance of the three-loop diagrams shown in Fig.~\ref{fig:higher-order-diagrams} (b), we examine the contribution from the isoscalar pion-nucleon vertex. In this case the contribution is given by 
%\begin{equation} 
%\Pi_{NN}(\omega)= \frac{T^{\pm}(\omega)}{280 \pi^6}  \Big[(C_S-3C_T)^2(11/2-\ln{2})(\kFp^7+\kFn^7)
%+(C_S^2+3C_T^2) \left(\kFp^7 F(\kFn/\kFp)+\kFn^7  F(\kFp/\kFn)\right)\Big]\,, 
%\label{eq:Pi_NN} 
%\end{equation} 
%where 
%\begin{equation}
%\label{eq:F-fcn}
%F(x) = \frac{x}{4}(15+33x^2-4x^4)+ \frac{1}{8}(15-42x^2+35x^4)\ln\left[\frac{|1-x|}{1+x}\right] +x^7 \ln\left[\frac{x^2}{|1-x^2|}\right]\,. 
%\end{equation}
\begin{equation} 
\Pi_{NN}(\omega,k_n,k_p)= \frac{T^{+}(\omega)}{280 \pi^6}  \Big[(C_S-3C_T)^2(11/2-\ln{2})(k_p^7+k_n^7)
+(C_S^2+3C_T^2) \left(k_p^7 \, F(k_n/k_p)+k_n^7 \, F(k_p/k_n)\right)\Big]\,, 
\label{eq:Pi_NN} 
\end{equation} 
where 
\begin{equation}
\label{eq:F-fcn}
F(x) = \frac{x}{4}(15+33x^2-4x^4)+ \frac{1}{8}(15-42x^2+35x^4)\ln \frac{|1-x|}{1+x} +x^7 \ln  \frac{x^2}{|1-x^2|} \,. 
\end{equation}
At low-density $\Pi_{NN}(\omega)$ is small compared to lower-order diagrams due to its higher $k_F$ dependence and can safely be neglected. Even at $n = 2 n_{\text{sat}}$, the highest density we consider, $\Pi_{NN}(\omega)$ is small compared to $\Pi_{ld}(\omega)$. Taking $C_S, C_T \sim (4\pi/M \mpi)$ (so KSW scaling), $\omega=200$ MeV, and $n = 2 n_{\text{sat}}$, one finds that $| \Pi_{ld}(\omega)/\Pi_{NN}(\omega) | \simeq  4$. Furthermore, at twice nuclear density, the uncertainties due to the $\Pi_{X_2}(\omega)$ diagrams are much larger than those due to $\Pi_{NN}(\omega)$. For these reasons, we neglect $\Pi_{NN}(\omega)$ in our analysis.

We conclude this section by recording once and for all the total self-energy of the negatively charged pion 
\begin{align}
\label{eq:full-self-energy}
    \Pi(\omega,k_n,k_p) &= \Pi_{ld}(\omega,k_n,k_p) + \Pi_{ds}(\omega,k_n,k_p) + \Pi_{cor}(\omega,k_n,k_p) + \xi(\omega) \Big( \Pi_{ph}(\omega,k_n,k_p) +  \Pi_{pw} (\omega,k_n,k_p)\Big) \nonumber\\
    &+ \sum_{X_2} \Pi_{X_2}(\omega,k_n,k_p),
\end{align}
where $\xi(\omega)$, given Eq. \ref{eq:enchancement}, is the enhancement factor arising from NN interactions. In the following results, uncertainty bands are obtained by varying $-1 \leq x_2 \leq 1$, subject to constraints imposed by pionic atoms that we discuss in section~\ref{symnuc}.

\section{A Mean Field Model for NN Interactions in Asymmetric Matter}\label{nucleonnucleon}
The energy difference between neutrons and protons can be large in dense matter. As already discussed in section \ref{subsec:2nucleon}, including this energy difference through the factor $\xi$ defined in Eq.~\ref{eq:enchancement} in calculating the pion self-energy is important because it alters energy denominators in diagrams with intermediate neutron-proton particle-hole states. In simple mean-field models, $\xi$ can be characterized by just two key parameters: the nuclear symmetry energy and the effective mass of the nucleons. In  general,  the strength of phenomenological, short-range interactions in mean field models is chosen to reproduce nuclear masses and bulk properties of matter, such as its energy density, pressure, and susceptibilities. A common feature of mean-field models is the modification of the single-particle nucleon energies due to their coupling to the mean field generated by other nucleons in the medium. In a large class of these models, the neutron and proton energies are given by
 \begin{align}
 E_n(p)  = {p^2\over 2M^*_n} + \Sigma_n,\, \qquad  
 E_p(p)  = {p^2\over 2M^*_p} + \Sigma_p\,.
 \label{eqn:n_disp}
 \end{align}
 Here, $\Sigma_{n,p}$  are the mean-field energy shifts, and $M^*_{n,p}$ are the effective masses of neutrons and protons in the medium. The mean field energy shift and the effective masses depend on the baryon density and the isospin asymmetry. For simplicity, we neglect the difference between neutron and proton effective masses and assume that $M^*_n=M^*_p=M^*$. 
 
Interestingly, in beta-equilibrated neutron star matter, $\xi(\omega)$ can be simplified by expressing $\Sigma_n - \Sigma_p$ in terms of the nuclear symmetry energy and the nucleon effective mass. This is accomplished by first noting that the isospin chemical potential $\hat\mu$, and the proton, neutron, and electron chemical potentials are related through
\begin{equation}
    \hat{\mu} = \mu_e = \mu_n-\mu_p~,
\end{equation}
and that furthermore $\mu_e=4 S(n_B)(1-2x_p)$, where $S(n_B)$ is the nuclear symmetry energy and $x_p$ the proton fraction. Substituting these relations into $\xi(\omega)$ yields
\begin{equation}
\xi(\omega) = \frac{\omega}{\omega - 4 S(n_B)(1-2x_p) +(k_n^2 - k_p^2)/2M^*} ~, 
\end{equation}
where $k_n=(3\pi^2 n_B(1-x_p))^{1/3}$ and $k_p=(3\pi^2 n_Bx_p)^{1/3}$ are the neutron and proton Fermi momenta. As stated previously, $\xi(\omega)$, when large, can change the importance of Feynman diagrams. To estimate this, note that if a pion is produced in the medium, then its frequency is forced to $\omega = \mu_e$. Making this substitution into $\xi(\omega)$ one finds that the diagram labelled $(ph)$, which originally contributed at ${\cal O}(k_f^5 \mu_e)$ now contributes at ${\cal O}(k_f^3 \mu_e^2)$; the diagram labelled $(pw)$, which originally contributed at ${\cal O}(k_f^6 \mu_e)$ now contributes at ${\cal O}(k_f^4 \mu_e^2)$; and the diagram labelled $(cd)$, which originally contributed at ${\cal O}(k_f^8 \mu_e)$ now contributes at  ${\cal O}(k_f^6 \mu_e^2)$. Crucially, in neutron-rich matter, these enhanced diagrams are attractive for positive frequencies. 

Through its appearance in the enhancement factor $\xi(\omega)$, the nuclear symmetry energy $S(n_B)$ plays an important role in our calculation. The nuclear symmetry energy is defined through the difference
\begin{equation}
S(n_B)= E(n_B,x_p=0)-E(n_B,x_p=1/2)\,,
\end{equation} 
where $E(n_B,x_p)$ is the energy per baryon at baryon density $n_B$ and proton fraction $x$. Microscopic calculations and fits to phenomenological models indicate that the energy per particle at arbitrary proton fraction is well approximated by 
\begin{equation}
E(n_B,x) \approx E(n_B,x_p=1/2) + S(n_B) (1-2x_p)\,, 
\end{equation}
since higher order terms in the expansion are small, even for $x_p \ll 1/2$ \cite{Li:2008gp}. In this case, the electron chemical potential in neutron star matter is 
\begin{equation}
\mu_e(n_B) = 4 S(n_B) (1-2 x_p) \,.     
\label{eq:mue}
\end{equation}
Recently, there has been much interest in determining the density dependence of the symmetry energy. However, despite progress in both theory and experiment, this dependence remains poorly known at densities reached in neutron stars. In the vicinity of nuclear saturation density, $S(n_B)$ impacts nuclear structure. Nuclear masses, measurements of the neutron-skin thickness, and the electric-dipole polarizability of neutron-rich nuclei such as $^{208}$Pb  provide useful constraints on $S_0=S(\nsat)$. Its density dependence is characterized by the slope parameter $L=3 n_B \left(dS(n_B)/dn_B\right)$ at $n_B=\nsat$  \cite{Steiner:2004fi}. 
Until recently, experiments, combined with theoretical models, suggested the  empirical range $ S_0= 32 \pm 2$ MeV and  $ L = 50 \pm 15$ MeV. Theoretical calculations using nucleon-nucleon interactions determined by \ChiEFT predict $S_0$ and $L$ compatible with this empirical range. For example, a recent calculation that combines many-body perturbation theory (MBPT) and Bayesian estimates for the truncation errors predicts $S_0=31.7\pm 1.1$ MeV and $L=59.8\pm 4.1$ MeV \cite{Drischler:2020hwi}. However, the recent measurement of the neutron-skin thickness of  $^{208}$Pb using parity-violating electron scattering imply larger values: $ S_0= 38.1 \pm 4.7$ MeV, and  $ L = 106 \pm 37$ MeV \cite{PREX:2021umo,Reed:2021nqk}.

% \begin{figure}[b!]
% \begin{centering}
% \includegraphics[width=0.55\textwidth]{Figures/SymEnergy.pdf}  
% \end{centering}
% \caption{Experimental and theoretical constraints on the density dependence of the nuclear symmetry energy. Predictions of the parametrization in Eq.\ref{eq:SymE2} are shown for reference.}
% \label{fig:SymE}
% \end{figure}

The symmetry energy at higher density can be accessed in heavy-ion experiments but is not presently well-determined \cite{Horowitz:2014bja}. At $n_B \simeq 1.5~\nsat$, a recent analysis by Estee et al. of charged pion yields from intermediate-energy heavy-ion collisions suggests that $S(n_B\simeq 1.5 \nsat) = 52\pm 13$ MeV \cite{SRIT:2021gcy}. Earlier studies by Russoto et al. of heavy-ion collisions at GSI indicate that $S(n_B \approx 2 \nsat) = 50\pm 7$ MeV \cite{Russotto:2016ucm}.

Theoretical calculations of the equation of state using potentials derived from \ChiEFT also provide useful constraints on the symmetry energy in the region $\nsat<n_B<2 \nsat$. Quantum Monte Carlo calculations by Lonardoni et al., using local \ChiEFT potentials predict $S(1.5\nsat)\approx 37 \pm 5$ MeV and  $S(2\nsat)\approx 46 \pm 11$ MeV \cite{Lonardoni:2019ypg}. In Ref.\,\cite{Lim:2018bkq}, Lim and Holt use many-body perturbation theory (MBPT) to predict $S(2\nsat)\approx 49 \pm 12$ MeV, and Drischler et al., combine MBPT and Bayesian estimates of the \ChiEFT truncation errors (but neglect errors associated with low energy constants) to predict  $S(2\nsat)\approx 45 \pm 3$ MeV \cite{Drischler:2020hwi}. 
In this study, we adopt a simple ansatz for the density dependence of the symmetry energy 
\begin{equation}
\tilde{S}(n_B)= \frac{2 S_2u}{2+ u}   \,, 
\label{eq:SymE2}
\end{equation}
where $u=n_B/\nsat$. Although this simple ansatz depends on a single parameter $S_2=S(2\nsat)$, the symmetry energy at twice saturation density, we find that $S_2$ in the range $45-60$ MeV satisfies all existing constraints. The high value $S_2=60$ MeV is compatible with the large $S_0$ and $L$ predicted by PREX and is implied by the heavy-ion data, and the intermediate value $S_2\simeq 50$ MeV is compatible with the empirical range. The low value  $S_2= 45$ MeV reasonably agrees with the \ChiEFT predictions made in Ref.~\cite{Drischler:2020hwi}. 
%A comparison between the model $\tilde{S}(n_B)$ and the predictions discussed are shown in Fig. \ref{fig:SymE}.

\section{Results}
\label{results}
In this section, we present our results for the pion mass in symmetric nuclear matter and neutron-rich matter encountered in astrophysics. The discussion of the pion mass in symmetric matter will help illustrate the importance of $\pi \pi NNNN$ interactions and highlight the need for improved constraints on the associated LECs. Results for the pion mass in beta-equilibrated neutron matter also assess the role $\pi \pi NNNN$ interactions and the nuclear symmetry energy. We find evidence for a  collective mode with the quantum numbers of the $\pi^+$ and its energy is sensitive to the nuclear symmetry energy. In the following, the $\pi N$ scattering parameters are fixed to $\sigma_N = 60$ MeV and $\gamma = 2.6$ and for the mean-field model we take $S_2 = 50$ MeV, and assume the in-medium nucleon mass $M^*(u) = M(1-\alpha u)$,  with $\alpha = 0.07$.

\subsection{Pion Mass in Symmetric Nuclear Matter}\label{symnuc}

As a prelude to calculating the pion masses in neutron-rich matter, we investigate the pion mass in symmetric nuclear matter. The calculation simplifies in this case, and the analytic results provide insights into the convergence of the EFT expansion, assess the importance of the two-nucleon operators, and compare with earlier work in Refs.~\cite{Thorsson:1995rj,Kolomeitsev:2002gc, Meissner:2001gz, Oller_2009,Voskresensky:2022gts,Goda:2013npa}. For symmetric matter where $k_n=k_p=k_f$, several isospin-odd contributions to the pion self-energy vanish, and since the neutron and proton mean-field energies are equal, $\xi(\omega)=1$.  One finds that 
\begin{equation}
\label{eq:pi-sym}
\Pi_{\rm sym}(\omega,k_f)=\Pi_{ld}(\omega,k_f,k_f) + \Pi_{ds}(\omega,k_f,k_f) + \sum_{X_2} \Pi_{X_2}(\omega,k_f,k_f) \,.
\end{equation}
While $\Pi_{\rm cor}$ is non-zero in symmetric matter, it is numerically small compared to the other terms, and we therefore neglect it in the expression above. The double scattering diagram contains pions in the intermediate state and encodes quantum corrections beyond the mean field approximation. In the region where $Q^2=\omega^2-\mpi^2>0$, and $k_f \gg Q$ (i.e., small positive shifts to the pion mass), we find  
the following analytic approximation for the self-energy in symmetric matter,
\begin{equation}
\label{eq:Pi_sym}
    \Pi_{\rm sym}(\omega,k_f) \simeq -T^+(\omega) n_B + \frac{\omega^2 k_f^2}{(2\pi f_{\pi})^4} \Big[ 2k_f^2 + Q^2\Big( 2\ln{  {|Q| \over 2k_f} }-1) \Big) \Big] 
%    \left(1-\frac{Q^2}{2k_f^2}\left(1-\ln{\frac{Q^2}{4k_f^2}}\right)  \right)  
-\frac{3 n_B^2}{16 f_{\pi}^2}(D_2 m_{\pi}^2 + D_2' m_{\pi}^2 + E_2 \omega^2 + E_2' \omega^2) \,.
\end{equation}
The in-medium pion mass $m_\pi^*$ is obtained by solving $\omega^2 - m_{\pi}^2 - \Pi_{\text{sym}}(\omega) = 0$. Further,  isospin symmetry in symmetric matter implies that shift $\delta m_{\pi}^2={m_{\pi}^*}^2 - m_{\pi}^2$ is the same for $\pi^-$ and $\pi^+$ and is given by the following implicit equation:
%\begin{equation} %I have checked this equation; it agrees exactly with the numerical solution
%\label{eq:analytic-mass}
%    \frac{\delta m_{\pi}^2}{m_{\pi}^2} 
%    = 
%    -\frac{1}{Z}~\left(\frac{T^+(m_{\pi})n_B}{m_{\pi}^2} + {2 k_f^2 \over (2\pi f_\pi)^4} -\frac{3 n_B^2}{16 f_{\pi}^2}(D_2 + D_2' + E_2 + E_2')  \right)\,,
%    { 1 - \frac{\sigma_N n_B}{m_{\pi}^2 f_{\pi}^2}  -\frac{32 k_f^4}{(4\pi f_{\pi})^4} \Big[1+ \frac{m_{\pi}^2}{2k_f^2}\ln\big(\frac{{Q^*}^2}{4 k_f^2 e}\big) + \frac{{Q^*}^2}{2k_f^2}\ln\big(\frac{{Q^*}^2}{4 k_f^2 e}\big) \Big] + \frac{3 n_B^2}{16 f_{\pi}^2}(E_2 + E_2')  }~.
%\end{equation}
\begin{equation} %I have checked this equation; it agrees exactly with the numerical solution
\label{eq:analytic-mass}
    \frac{\delta m_{\pi}^2}{m_{\pi}^2} 
    = 
    \frac{\frac{-T^+(m_{\pi})n_B}{m_{\pi}^2} + {2 k_f^2 \over (2\pi f_\pi)^4} -\frac{3 n_B^2}{16 f_{\pi}^2}(D_2 + D_2' + E_2 + E_2')  }{1 -\frac{\sigma_N n_B}{m_{\pi}^2 f_{\pi}^2}  +\frac{k_f^2}{(2\pi f_{\pi})^4}
   \Big( \, m_\pi^2+\delta m_{\pi}^2 -2k_f^2- (m_\pi^2+\delta m_{\pi}^2) \ln{ {| \delta m_{\pi}^2 | \over 4k_f^2} } \, \Big) + \frac{3 n_B^2}{16 \fpi^2}(E_2 + E_2')   }
   \, .
\end{equation}
%where  \SR{Neill, please check}
%\begin{equation} %I have checked this analytic result agress with the numerical result.
%\label{eq:Zfactor} 
%    Z = 1 -\frac{\sigma_N n_B}{m_{\pi}^2 f_{\pi}^2}  +\frac{k_f^2}{(2\pi f_{\pi})^4}
%   \Big( \, m_\pi^2+\delta m_{\pi}^2 -2k_f^2- (m_\pi^2+\delta m_{\pi}^2) \ln{ \Big \lvert {\delta m_{\pi}^2\over 4k_f^2} \Big \rvert } \, \Big)
%    \Bigg[1-\frac{{Q^*}^2}{2k_f^2}+\frac{{Q^*}^2}{k_f^2}\ln \big(\frac{{Q^*}^2}{4k_f^2}\big) + \frac{m_{\pi}^2}{2k_f^2}\ln \big(\frac{{Q^*}^2}{4k_f^2}\big) \Bigg] + 
%+ \frac{3 n_B^2}{16 f_{\pi}^2}\big(E_2 +E_2' \big) ~.
%\end{equation}
%is the wavefunction renormalization factor defined through the inverse of pion propagator as $i Z (p^2-{m_{\pi}^*}^2)$ as $p^2 \rightarrow {m_{\pi}^*}^2$ and can be obtained through $Z = 1 - \frac{\partial \Pi}{\partial Q^2} \rvert_{\delta m_{\pi}^2}$. We note that our result for $Z$ linear in the density matches previous calculations \cite{Meissner:2001gz}. 
%At low density $\delta m_{\pi}^2 \propto T^+(m_{\pi}) k_f^3$, which implies that logarithmic correction to the pole position is small. 

%The width of the pion is \cite{Peskin:1995ev}
%\begin{equation}
%\label{eq:width}
%    \Gamma = -\frac{Z(m_{\pi}^*)^{-1}}{m_{\pi}^*}\text{Im }\Pi_{\text{sym}}(m_{\pi}^*)~,
%\end{equation}

\begin{figure*}[h]
    \includegraphics[height=3.5in]{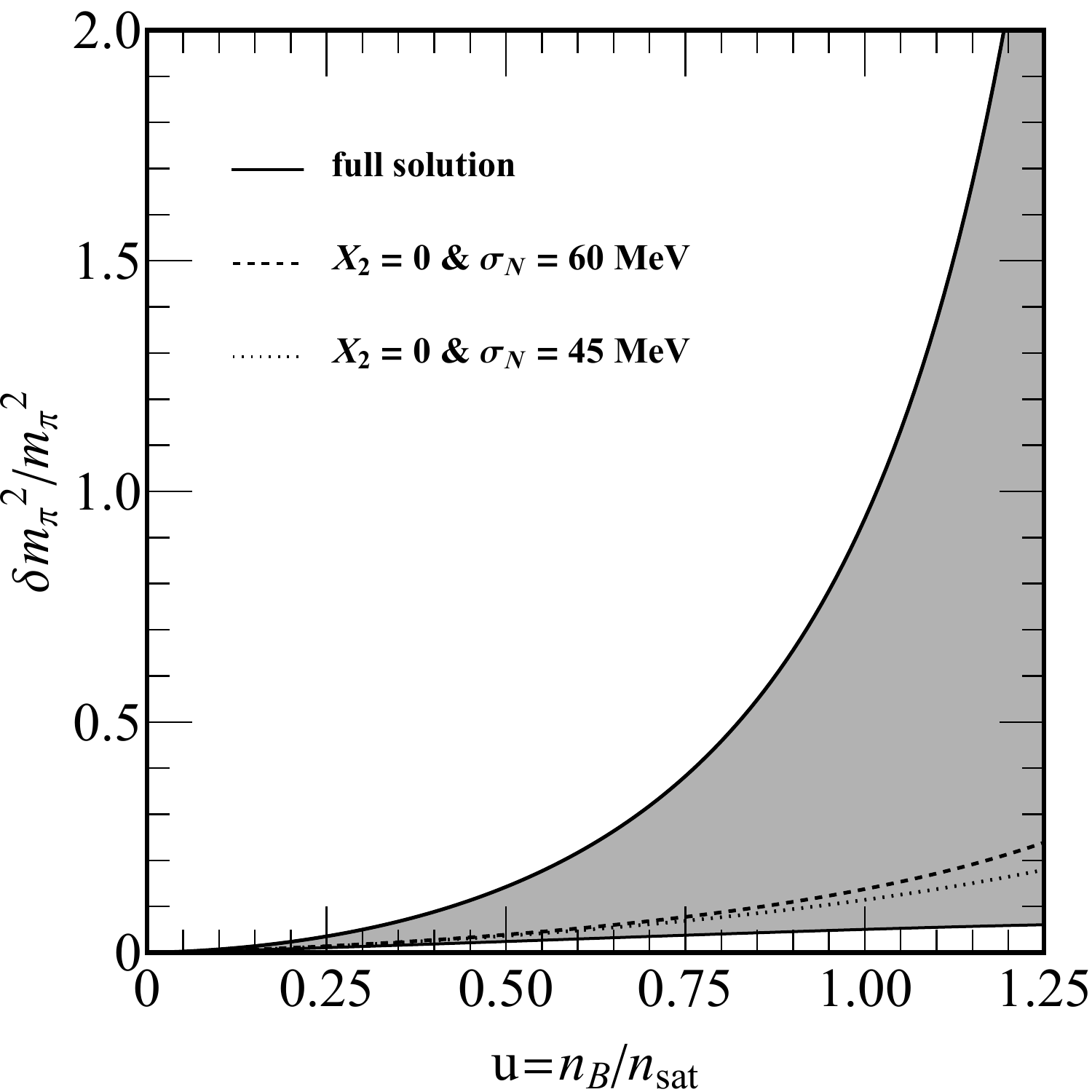}  
    \caption{Plot of $\delta m_{\pi}^2/m_{\pi}^2$ in symmetric matter. The error bands are generated by varying $x_2$ over the range of values allowed by pionic atoms constraints. The dashed and dotted curves on the left are obtained by setting all $X_2 = 0$ and varying the sigma term $\sigma_N$.} 
    %In both, dotted curves are obtained by taking all $x_2 = 0$ while the bands are generated by varying $-1 \leq x_2 \leq 1$.
\label{fig:mass-shift-sym-matter}
\end{figure*} 

%The central dotted curves in both panels are obtained by taking all $x_2=0$, while the light grey bands are generated by varying the four couplings over the range $-1\leq x_2 \leq 1$. 
Work on pionic atoms (for reviews, see \cite{Batty:1997zp,Friedman:2007zza}) based on the optical potential model has provided broad constraints on the effective mass of the pion in symmetric matter at sub-saturation density. These analyses assume a specific dependence on density motivated by phenomenological models for the optical potential denoted by $V_{\rm opt}(n_n,n_p)$ and employ measurements of the energy levels and widths of pion atoms to constrain the model parameters \cite{Friedman:2002ix}. The self-energy defined in Eq.~\ref{eq:Pi_sym} is related to the optical potential, and the sum of the LECs  $(D_2+D'_2+E_2+E'_2)$ and $(E_2+E'_2)$ can be constrained by developing a new optical potential that is faithful to the energy and density dependence predicted by Eq.~\ref{eq:Pi_sym} and employing it refit the pionic atom data. Performing such fits to constrain the two-pion-two-nucleon LECs is beyond the scope of this work. Instead, we adopt a simple and approximate procedure to implement constraints from pionic atom spectroscopy. In particular, we use the models of \cite{Friedman:2002ix} to bound the pion mass at $n = 2/3 n_0$. We find an allowed mass range of $0.05 \leq \delta \mpi^2/\mpi^2 \leq 0.35$, then use this to bound the $X_2$ coefficients, whose range produces the gray band in Fig. \ref{fig:mass-shift-sym-matter}. We find that this procedure is roughly compatible with the estimate that $|x_2| \leq 1$; however, further work is needed to implement pionic atom constraints properly. In particular, we have not considered correlations between the sigma term and the $X_2$ operators. Presumably, a proper accounting will increase the uncertainty in the pion mass.
%\textcolor{red}{We require that $0.08 \leq \delta \mpi^2/\mpi^2 \leq 0.16$ at $n_B = 2/3 n_0$. This bound produces a range of allowed $x_2$, which produce the gray bands in Fig. \ref{fig:mass-shift-sym-matter}. We find that this demand is compatible with the estimate that $|x_2| \leq 1$, however further work is needed to implement pionic atom constraints properly.}
%Using the conventional model for $V_{\rm opt}$ described in  \cite{Friedman:2002ix}, we derive approximate bounds on $\delta m_\pi^2$ for $0.5 ~\nsat <n_B \nsat$, which is indicated by the dark shaded region in Fig. \ref{fig:mass-shift-sym-matter}. The thick solid and thick dashed curves, obtained by varying $x_2$ in the range $-1 < x_2 < 1$ subject to the constraints imposed by the dark-shaded regions, depict the lower and upper bounds of our predictions for the mass shifts in symmetric nuclear matter. 
The preceding analysis and the results in Fig. \ref{fig:mass-shift-sym-matter} provide several insights:  
\begin{itemize}
    \item The leading correction to the pion mass given by $-T^+(m_{\pi})n_B$ is linear in the baryon density and depends only on the on-shell pion-nucleon scattering amplitude. This general form is expected from low-density theorems. However, since $T^+(m_{\pi})\simeq 0$, the $\mathcal{O}(k_f^4)$ contribution to the pion mass and sub-leading corrections play a more important role. For example, the contribution from the $\mathcal{O}(k_f^6)$ $X_2$ operators can impact the density dependence of the pion mass and the chiral condensate even at modest densities. 
    \item  The $X_2$ operators would need to be included in the analysis of pionic atoms.  From Eq.~\ref{eq:analytic-mass}, we can deduce the contribution from the $X_2$ operators is of similar size at a density $n_B \simeq 0.6 ~\nsat$, which is expected to the average density probed in pionic atoms \cite{Friedman:2002sys}. In this context, the phenomenological optical potential models fit to pionic atom data suggests $0.05 \leq \delta \mpi^2/\mpi^2 \leq 0.35$ at $n = 2/3 n_0$,. As a first step, we used this to bound the $X_2$ coefficients by assuming that the other LECs are fixed and $\sigma_{N}=60$ MeV. However, since the $X_2$ operators make a contribution that is comparable to that induced by the $\sigma_{N}$-term in the denominator of Eq.~\ref{eq:analytic-mass}, pionic atoms constraints on $\sigma_{N}$ will likely need to be revised. 
    \item In earlier analysis that neglects the two-nucleon contributions, the positive shift of the pion mass at finite density is understood within \ChiPT as arising due to a combination of effects which include the energy dependence of pion-nucleon scattering amplitude, the double scattering contribution and the corrections due to renormalization of the wave function at $\mathcal{O}(k_f^6)$ and $\mathcal{O}(k_f^7)$\cite{Kolomeitsev:2002gc,WEISE200198,Friedman:2002ix,Friedman:2002sys}. In these previous analyses and our accounting, the $\mathcal{O}(k_f^6)$ terms are incomplete; nucleon-nucleon interactions and low energy constants, as well as the $X_2$ two-nucleon currents, all contribute at $\mathcal{O}(k_f^6)$. The only firm conclusion we can draw at this stage is that a more detailed \ChiPT analysis that includes the $X_2$ operators and NN interactions is needed to interpret data from pionic atoms. Such an analysis can provide useful constraints on the sum $D_2 + D'_2 + E_2+ E'_2$.  
    \item The rapid increase of the two-nucleon contribution with density indicates that predictions relating to pion condensation in symmetric nuclear matter at $n_B \gtrsim \nsat$, such as those discussed in  \cite{Voskresensky:2022gts} which neglected two-nucleon contributions, need to be revisited. The growth of the multi-nucleon contribution to the pion dispersion relations seen in Fig. \ref{fig:mass-shift-sym-matter} suggests the convergence of \ChiPT becomes an issue for $n_B \gtrsim 1.5~\nsat$.  
\end{itemize}

\subsection{Pion Mass in Neutron-rich matter}
\label{sec:neutron-rich}
We next consider the properties of charged pions in beta-equilibrated nuclear matter, where $\hat{\mu} = \mu_e = \mu_n-\mu_p$. The poles of the propagator, or the zeros of the inverse propagator at zero momentum, $\omega^2 - m_\pi^2 - \Pi(\omega, k_n,k_p)=0 $, correspond to real excitations in the medium. The energy and charge associated with these excitations are determined by examining the residue at the pole. When the mean field splitting of the neutron and proton energies is neglected, the propagator has two poles, as expected in the vacuum. The pole at positive frequency has a positive residue. Since we are examining the two-point function of a \emph{negatively} charged field defined in Eq.~\ref{eq:piminus_twopoint}, this implies that the charge associated with the pole at $\omega_+$  is negative and its effective mass $m^*_{\pi^-}=\omega_+$. The pole at negative frequency denoted by $\omega_-$ has a negative residue. It thereby corresponds to the $\pi^+$ state with effective mass $m^*_{\pi^+}=-\omega_-$. When the mean field splitting between the neutron and proton energy is included, as discussed below, the pion propagator contains an additional pole. This has a negative residue and is identified as a positively charged collective state associated with the particle-hole excitation spectrum. This mode is labeled as $\pi_s^+$ to be consistent with the notation used in earlier work to describe a similar mode at finite momentum \cite{Au:1974mbl}. 

\begin{figure*}[h]
    \centering
       \includegraphics[height=3.5in]{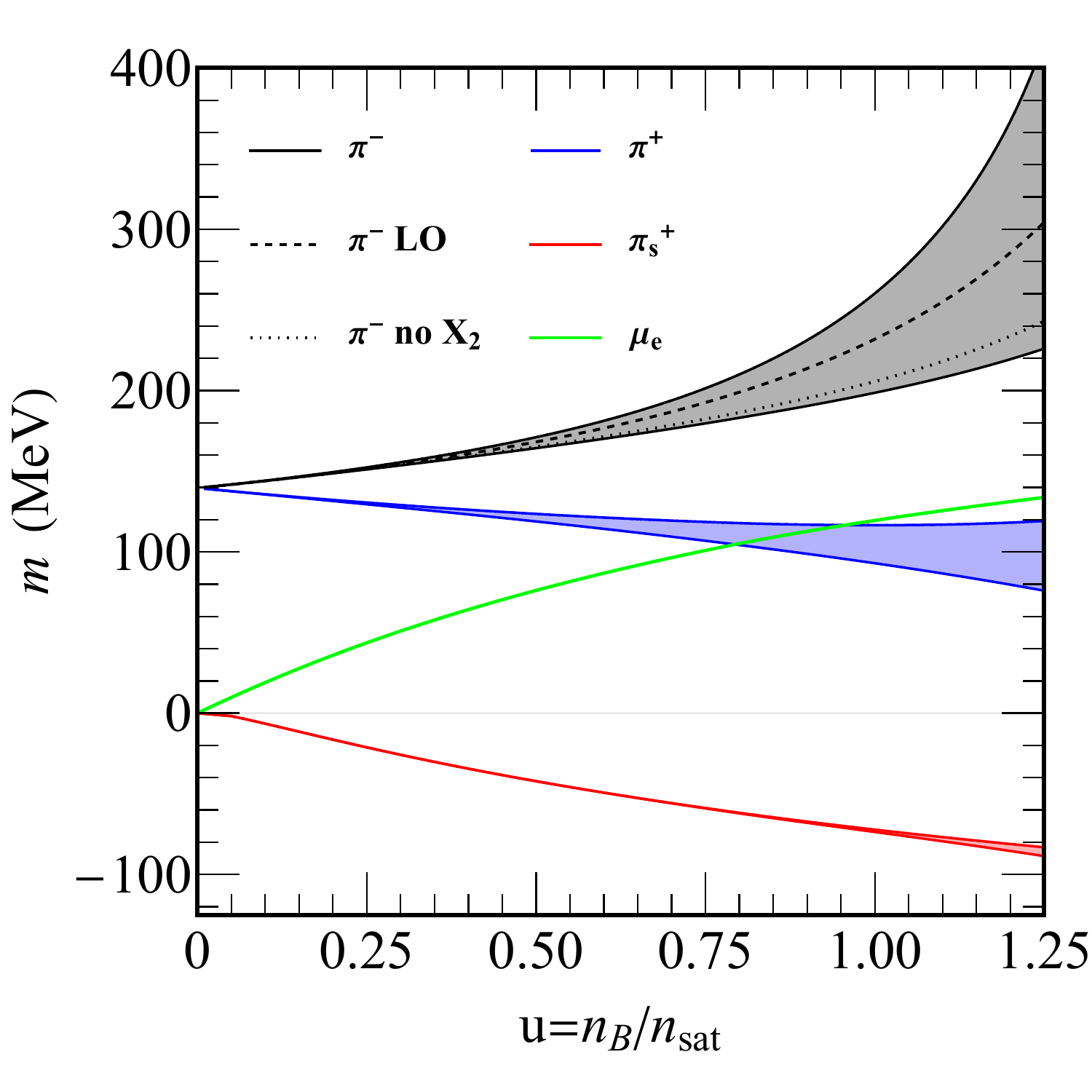}
        \caption{Masses of the charged pions and the pion-like collective excitation in beta-equilibrated, neutron-rich matter from the full self-energy Eq. \ref{eq:full-self-energy}.  The error bands in both are generated by varying the $x_2$ over the range allowed by pionic atoms constraints. The dotted and dashed black lines show the contributions to the pion mass from the leading-order self-energy and the self-energy with all $X_2 = 0$ .}
\label{fig:energies}
\end{figure*}
Our results are summarized in Fig. \ref{fig:energies}, where the 
$\pi^-, \pi^+, \pi^+_s$ masses are plotted in black, blue, and red, respectively, and the electron chemical potential given in Eq. \ref{eq:mue} is plotted in green. The leading-order s-wave WT term dominates at low density, producing a linear increase (decrease) in the $\pi^-$($\pi^+$) mass. 
%The masses of the $\pi^-$ and $\pi^+$ obtained using Eq.~\ref{eq:full-self-energy} by neglecting two-nucleons contributions, i.e., $x_2=0$, and for the parameters specified earlier are shown as the black and blue dotted curves Fig. \ref{fig:energies}. 
The nearly linear behavior of the black and blue curves implies that corrections due to $\Pi_{ds}$, $\Pi_{ph}$,$\Pi_{pw}$ are relatively small. This can also be seen by comparing the sold black band to the dashed black curve, which is the pion mass that results from the leading order self-energy. Furthermore, by comparing the dotted curve - which is obtained from the full self-energy but with all $X_2$ set to zero - to the leading-order result, it's found that higher order corrections to the pion self-energy slightly lower the pion mass.
%The green curve shows the electron chemical potential $\mu_e$ for the specific model of the symmetry energy described by Eq.~\ref{eq:SymE2} with $S_2=50$ MeV. 

The finite energy difference between neutrons and protons produced by mean-field effects generates a new collective mode with an energy close to that associated with the single neutron-hole proton-particle state. This new pole in the pion two-point function arises because $\xi(\omega)$ is large when $\omega \simeq \Sigma_n-\Sigma_p$ and this enhances the $\Pi_{ph}$ and $\Pi_{pw}$ contributions to the total self-energy in Eq.~\ref{eq:full-self-energy}.  The energy of this collective excitation is shown by the red curve in Fig. \ref{fig:energies}. The collective state has the quantum numbers of the $\pi^+$ and arises as a coherent proton-particle--neutron-hole excitation. 

We note that the negative energy of the $\pi_s^+$ does not imply instability. The large electron chemical potential ensures that energy associated with producing $\pi_s^+$ through the reactions $p\rightarrow n \pi_s^+ $, given by $\delta \omega_s=\mu_n-\mu_p+\omega_{\pi_s^+}=\omega_{\pi_s^+}+\mu_e$ is positive. The ground state is generally unstable when $\delta\omega_i=m^*_i- \mu_i <0$, where $\mu_i$ is the chemical potential associated with the conserved charge labeled $i$, and $m^*_i$ is the effective mass of the particle in the ground state. In charge neutral neutron-rich matter in beta-equilibrium, the electric charge chemical potential $\mu_Q=-\mu_e$, where  $\mu_e=\mu_n-\mu_p>0$.

The error bands are calculated by including the contribution of the $X_2$ operators as described in the last section.
%In our conservative approach, we allow all four dimensionless coefficients $d_2,e_2$ and $d'_2,e'_2$ to take values between $-1$ and $1$. 
%
%The rapid increase of width of the error band with density is noteworthy. 
%
%Already at $n \simeq \nsat $ the uncertainty in the mass of the $\pi^-$ is of the order of $100$ MeV.  
In neutron-rich matter, the contribution from $\Pi_{D_2}$ and $\Pi_{E_2}$ are the most important $X_2$ operators, and the latter, which is proportional to ${m_{\pi^-}^*}^2$, increases the associated uncertainty. 
%If the pionic atom constraint, we discussed earlier, on the sum $d_2+e_2+d'_2+e'_2$ is imposed, the width of the bands decreases, \SR{does this preclude pion condensation?}. In the extreme low-mass case, intermediate-density s-wave pion condensation is not definitively ruled out, as shown in Fig. \ref{fig:energies}.  
The uncertainty associated with the $\pi^+$ mode is smaller because the leading WT contribution is attractive in this case. This lowers ${m_{\pi^+}^*}^2$ and the energy dependent contribution $\Pi_{E_2}$ is correspondingly smaller. The energy of the $\pi_s^+$ mode is not sensitive to the $X_2$ contributions. However, the energy of this mode depends sensitively on the nuclear symmetry energy, and this uncertainty is not included in the bands depicted in Fig.~\ref{fig:energies}.

\section{Conclusions}
\label{conclusions}
We have constructed a model for the pion self-energy using a combination of heavy baryon chiral perturbation theory for pion-nucleon interactions and a simple phenomenological model for nucleon-nucleon interactions. Within the single nucleon sector, we include all diagrams up to $\mathcal{O}(q^6)$ in the low-momentum expansion. We augment these (known) results by accounting for multi-nucleon interactions in two ways. First, we account for nucleon-nucleon interactions by introducing a self-energy into nucleon propagators. Such a modification can be motivated by examining the effect of leading order NN interactions into nucleon lines. The nucleon self-energy is constructed from a mean-field model which incorporates known constraints on the symmetry energy of nuclear matter. We construct a one-parameter model $\tilde{S}(n_B)$ for the symmetry energy, which approximately reproduces all known constraints. Second, and perhaps more importantly, we include pion-multi-nucleon operators and find that their contribution to the pion mass increases rapidly with density if they are of natural size, and the associated uncertainty is large because the relevant LECs are poorly constrained. Interestingly, the relative importance of the leading two-pion-two-nucleon operators is especially significant in nuclear matter with a nearly equal number of neutrons and protons. In this case, the contribution to the pion mass at the leading order in the density is small because the isoscalar scattering amplitude nearly vanishes at the threshold. Including these operators in the reanalysis of pionic atoms and pion-nucleus scattering could provide valuable constraints on some combinations of the LECs associated with the two-pion-two-nucleon operators.

In neutron-rich matter, our results for the charged pion masses provide the following insights. First, the contribution from the higher-order ($\mathcal{O}(q^5)$, and $\mathcal{O}(q^6)$) diagrams arising from pion couplings to single nucleons are modest. The WT term ($\mathcal{O}(q^4)$) makes the dominant contribution to $m_{\pi^-}$ over the entire range of densities considered. The net effect of the $\mathcal{O}(q^5)$, and $\mathcal{O}(q^6)$ diagrams that neglect the $X_2$ contributions is to lower the pion mass by about ten percent. At $n_B=\nsat$, the double scattering diagram $\Pi_{ds}$ is produces a positive mass shift of $\sim 5\%$, while  both $\Pi_{ph}$ and $\Pi_{pw}$ produce negative $~10\%, 7\%$ shifts, respectively. Despite the large uncertainity associated with contributions from the $X_2$ operators, pionic atom constraints  on the LECs ensure that $m_{\pi^-}$ continues to increase with density and disfavors $\pi^-$ condensation. The decrease of $m_{\pi^+}$ with density is nearly linear, and the higher order contributions from the two-loop and $X_2$ operators is modest even at the highest densities we considered.

Our prediction of a positively charged collective mode in the long-wavelength limit is new. The energy of this mode is negative, and its value is sensitive to the energy splitting of neutrons and protons in the medium, the one-loop particle-hole diagram, and the two-loop diagram that involves a p-wave interaction between pions and nucleons. The phenomenological implications of the low-energy collective mode and the large splitting between the masses of the charged pion excitations warrant further study. Interestingly, a negative energy spin-isospin collective mode was discussed in the context of neutrino reactions in the warm neutron-rich matter \cite{Shin:2023sei}. The relationship between the spin-isospin collective mode and $\pi^+_s$, and their role in neutrino production and absorption reactions and affect transport of heat and lepton number in supernovae and neutron star mergers warrants further investigation.

\section*{Acknowledgements} The work of S. R. and N. C. W. was supported by the U.S. DOE under Grant No. DE-FG02- 00ER41132. B. F. acknowledges support from the SciDAC Grant No. A18-0354-S002 (de-sc0018232).  We thank Paulo Bedaque, Gordon Baym, Vincenzo Cirigliano, Evgeny Epelbaum, David Kaplan, Bira van Kolck, Emanuele Mereghetti, Martin Savage, Thomas Schafer, Achim Schwenk,  Corbinian Wellenhofer, and Dima Voskresensky for helpful conversations. We also thank Avraham Gal and Eli Friedman for useful correspondence relating to pionic atoms. The work of S. R. was performed in part while attending the ``Exploring Extreme Matter in the Era of Multimessenger Astronomy: from the Cosmos to Quarks" workshop at the Aspen Center for Physics.  

\bibliography{pion}
\appendix

\section{Loop functions }
In this appendix, we give analytical expressions for  the pertinent functions $I(\omega; k_n,k_p), I(0,k_n,k_n), K(\omega; k_n, k_p)$ and $K(0;k_n,k_p)$ that appear in the contribution $\Pi_{pw}(\omega)$ to the pion selfenergy. In order to evaluate the Fermi sphere integrals in Eq.(17), three nontrivial integrations over the cosine of an inclined angle and two radii have to be performed. The angular integral leads to logarithms, and treating these as logarithms of absolute values effectively implements the principal-value prescription. One finds:
\begin{align}
    384\pi^4 I(\omega;k_n,k_p) = {32 \over 3}k_n^3k_p^3 +Q^2 L(\omega; k_n, k_p)\,,
\end{align}
with $Q = \sqrt{\omega^2-m_\pi^2}$ and  the function $L(\omega; k_n, k_p)$ is written in Eq.(13). The other loop functions read:

\begin{align}
    384\pi^4 I(0;k_n,k_n) = {32 \over 3}k_n^6-24 k_n^4m_\pi^2+4k_n^2m_\pi^4+32 k_n^3 m_\pi^3 \arctan{2k_n \over m_\pi}-m_\pi^4(12k_n^2+m_\pi^2)\ln\Big( 1 + {4k_n^2\over m_\pi^2}\Big)\,,
\end{align}
\begin{align}
    192\pi^4 K(\omega;k_n,k_p) =&\,\, 4k_n k_p(k_n^2-k_p^2)(k_n^2+k_p^2-5Q^2) +(k_n^2-k_p^2)\Big[(k_n^2-k_p^2)^2-6Q^2(k_n^2+k_p^2)-3Q^4\Big]  \nonumber \\ & \times \ln{|(k_n-k_p)^2-Q^2| \over |(k_n+k_p)^2-Q^2|}  + 8Q^3\bigg[ (k_n^3+k_p^3)  \ln{|k_n-k_p-Q| \over |k_n-k_p+Q|} +(k_n^3-k_p^3)  \ln{k_n+k_p+Q \over |k_n+k_p-Q|}\bigg] \,,
\end{align}\begin{align}
    192\pi^4 K(0;k_n,k_p) =&\,\, 4k_n k_p(k_n^2-k_p^2)(k_n^2+k_p^2+5m_\pi^2) +(k_n^2-k_p^2)\Big[3m_\pi^4
-6m_\pi^2 (k_n^2+k_p^2)-(k_n^2-k_p^2)^2\Big]  \nonumber \\ & \times \ln{(k_n+k_p)^2+m_\pi^2 \over (k_n-k_p)^2+m_\pi^2}  + 16m_\pi^3\bigg[ (k_n^3+k_p^3)  \arctan{k_n-k_p \over m_\pi}+(k_p^3-k_n^3)  \arctan{k_n+k_p \over m_\pi}\bigg] \,.
\end{align}

\end{document}